\documentclass[twocolumn,nofootinbib]{revtex4}

\usepackage{graphicx}

\begin{document}

\title{Gravitational waves from binary axionic black holes}

\author{J. A. de Freitas Pacheco$^1$, S. Carneiro$^{2,3}$, J. C. Fabris$^{4,5}$}

\affiliation{$^1$Universit\'e de Nice - Observatoire de la C\^ote d'Azur, 06304, Nice, France\\$^2$Instituto de F\'{\i}sica Gleb Wataghin, UNICAMP, 13083-970, Campinas, SP, Brazil\\$^3$Instituto de F\'{\i}sica, Universidade Federal da Bahia, 40210-340, Salvador, Bahia, Brazil\\
$^4$N\'ucleo Cosmo-ufes \& Departamento de F\'{\i}sica, UFES, 29075-910, Vit\'oria, ES, Brazil\\$^5$National Research Nuclear University MEPhI, 115409, Moscow, Russia}

\date{\today}

\begin{abstract}

In a recent paper we have shown that a minimally coupled, self-interacting scalar field of mass $m$ can form black holes 
of mass $M=\sqrt{3}/(4m)$ (in Planck units). If dark matter is composed by axions, they can 
form miniclusters that for QCD axions have masses below this value. In this work it is shown that for a scenario in which the axion mass 
depends on the temperature as $m \propto T^{-6}$, minicluster masses above $0.32\,M_\odot$, corresponding to an axion mass of $3\times 10^{-10}$ eV, exceed $M$ and can 
collapse into black holes. If a fraction of these black holes is in binary systems, gravitational waves emitted during the inspiral
phase could be detected by advanced interferometers like LIGO or VIRGO and by the planned Einstein Telescope. For a detection rate
of one event per year, the lower limits on the binary fraction are $10^{-4}$ and $10^{-6}$ for LIGO and Einstein Telescope respectively.

\end{abstract}
 
\maketitle

\section{Introduction}

\begin{figure}
\begin{center}
 \includegraphics[width=0.35\textwidth]{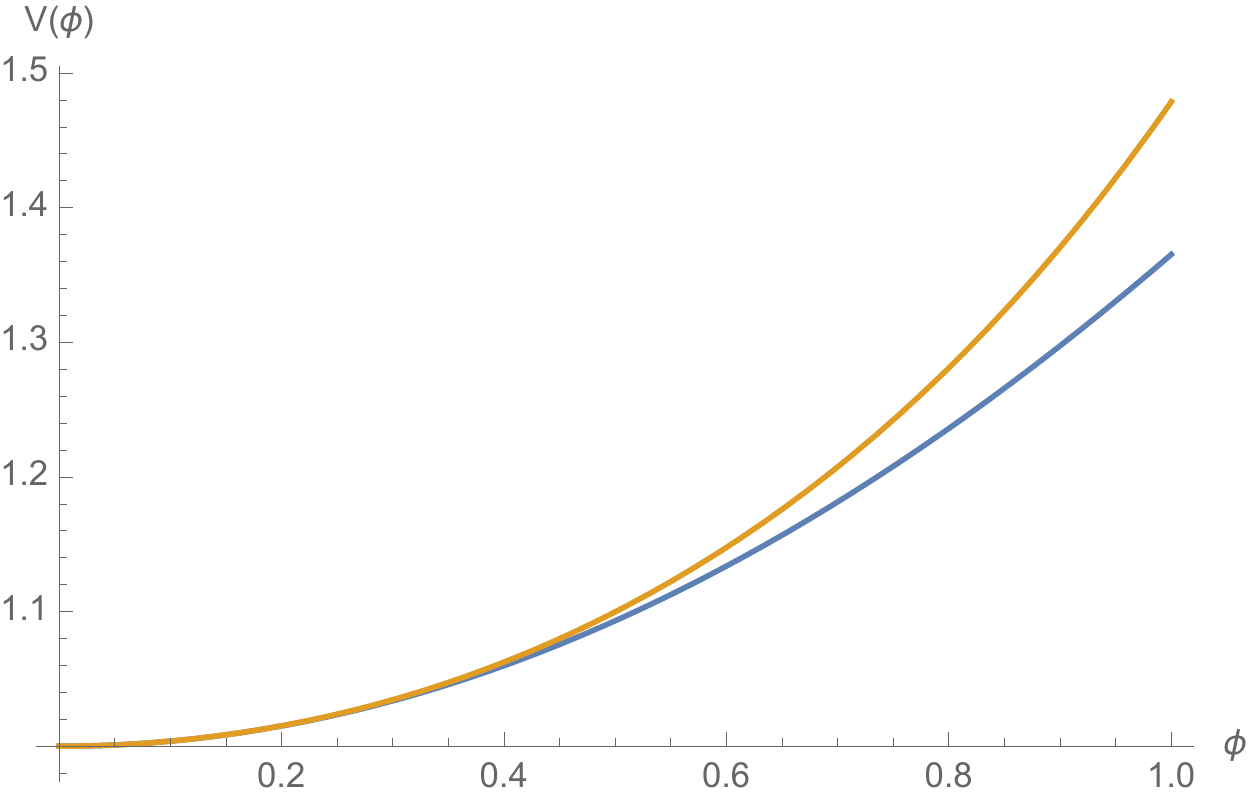} 
\end{center}
\caption{Comparison between the axionic potential \cite{russos3} (blue) and the hyperbolic potential used in \cite{EPJC} (yellow), for 
the same initial conditions.}
 \label{fig1}
\end{figure}

Still hypothetical particles, with masses estimated to be in the range $m \sim 10^{-22}$ eV up to $10^{-3}$ eV and with a small interaction cross section, axions 
are up to date the most reliable theoretical explanation for the CP problem of strong interactions \cite{CP}. In 
a recent paper \cite{EPJC} we have shown that scalar fields may, in general, suffer gravitational collapse, producing a central singularity 
(see also \cite{zeldovich,russos,russos2}) and the black hole formed under this process
has a mass $M=\sqrt{3} M_P^2/(4m)$. This result was derived from an exact solution of the Einstein-Klein-Gordon equations in 
spherical symmetry, assuming that the scalar field was minimally coupled to gravity. A particular hyperbolic self-interacting potential was adopted, which 
mimics a free field potential during the first stages of the collapse. For different potentials the collapse (if it occurs) will follow a different evolutionary 
behavior. Nevertheless, since an initially free field can trigger the collapse, we expect that this happens in general, except for the particular case of a 
strongly repulsive interaction. In fact, both a $\lambda\phi^4$ potential or the axion potential given in \cite{russos3} are less repulsive than the hyperbolic potential 
adopted in \cite{EPJC}. This can be shown in Fig. \ref{fig1} where the axion potential of \cite{russos3} is compared with the potential of  
\cite{EPJC} for identical initial conditions. On the other hand, Kaup in a seminal paper \cite{Kaup} solved numerically for the first time the general relativistic 
Klein-Gordon equation, deriving the eigenstates for spherical symmetry. He concluded that in the free field case there is a maximum mass given by 
$M_{\text{max}}=0.633\, M_P^2/m$ (the so-called Kaup limit), above which the system becomes unstable. 
Notice that the Kaup limit differs slightly in the numerical factor as compared to the mass $M$ mentioned above, probably a consequence of being the result
of a series of static equilibrium solutions while the result by \cite{EPJC} was obtained from a dynamic study of the black hole formation. The inclusion of a self-interaction with potential $V(\phi)=(1/4)\lambda\phi^4$ increases dramatically the maximum mass to $M_{\text{max}}=0.127\lambda^{1/2}(M_P^3/m^2)$ (see, for instance, \cite{Colpi}). On the other hand, attractive self-interactions lower the maximum mass below the Kaup mass, leading to an instability to axion-nova before the collapse to a black hole \cite{russos3}. It should also be emphasised that the maximum mass observed in the diagram ``mass versus number of particles", often called the ``critical mass" in the literature, is not necessarily the value above which the system collapses as it was shown by Gleiser \cite{Gleiser}. This is also true in the case of dense axion stars, which are unstable for masses above this limit and only collapse to black holes for masses still higher \cite{russos3}.

On the other hand, axions are good candidates to be cold dark matter particles present in the standard cosmological model \cite{cosmology}, because they interact very 
weakly either with baryonic matter or radiation. Axions are produced non-thermally and are the consequence of a spontaneously broken global symmetry, known as a
Peccei-Quinn symmetry, which occurs when the temperature of the universe drops below the symmetry breaking scale.
If dark matter is composed, at least partially by axions, it is not possible to exclude that mini-axion stars (dubbed miniclusters in the literature) 
could have been formed from density  fluctuations present in the epoch of symmetry breaking. In fact, the phenomenology of the minicluster formation is determined by the 
cosmic epoch during which the symmetry breaking occurs. Hogan \& Rees \cite{rees} considered a density contrast of the order 
of the unity and that the typical mass of a mini-axion star corresponds to the mass of all axions inside the horizon at $T \sim 100$ MeV. In this case,
the masses of these miniclusters would be about $10^{-6}~M_\odot$. However, Kolb \& Tkachev \cite{russos} found that oscillations
of the axion potential are responsible for non-linear effects controlling the density of mini-axion stars. The mass scale of these objects is
fixed as before by the total mass in axions within the horizon but at $T \sim 1$ GeV, when the axion mass is of the order of $~10^{-9}$ eV. These 
effects reduce the mini-axion star mass to about $10^{-12}~M_\odot$. These masses are below the critical mass
expected to trigger the gravitational collapse. Recently, Fairbairn et al. \cite{fairbairn} revisited the formation of miniclusters, considering
different scenarios for the evolution of the axion mass. Here, their results will be used to show that miniclusters of non-QCD axions having masses 
above $0.32~M_\odot$, corresponding to an axion mass of $\sim 3\times 10^{-10}$ eV can collapse and form black holes. In \cite{fairbairn}, the 
Press-Schechter formalism was used to compute the mass function of dark halos formed during the hierarchical structure process in which the
miniclusters are seeds. Presently, we cannot exclude the possibility that a small fraction of collapsed miniclusters could constitute binary
systems formed during successive merger episodes that led to the assembly of dark halos. In this work we will explore this possibility, aiming to constrain the putative
fraction of binaries formed during merger events by using the gravitational wave signal emitted during the coalescence of these systems.
We will make predictions for the advanced-LIGO \cite{pitkin} and for the planned Einstein Telescope (\cite{ET}) laser interferometers. The paper is
organized as follows: in Section II we discuss the critical mass of miniclusters able to collapse, in Section III 
a simple evolutionary model for the pairs is discussed, in Section IV we give a detailed discussion of the detection of the events 
and in Section V we present our final considerations.

\section{Critical Mass}

The formation of compact structures constituted by axions or axion-like particles (ALP) is plagued by the absence of an effective cooling mechanism 
\cite{fukugita}. In the case of configurations involving scalar fields, different investigations indicate that these systems may relaxe through
the emission of bursts of particles \cite{seidel,guzman}, a process known as gravitational cooling. Coherent oscillations of the scalar field
may also help the relaxation process that leads to the formation of a compact boson star \cite{tkachev}. Another path to form compact axion-like 
structures was investigated in \cite{shive}, where dark mater was assumed to be composed by a non-relativistic bosonic condensate in which the uncertainty 
principle balances gravity for scales less than the Jeans length. The high resolution cosmological simulations performed with this dark matter model 
with $m \sim 10^{-22}$ eV indicate that the resulting large structure is indistinguishable 
from cold dark matter, but ``solitonic'' compact structures are generally formed in the core of galaxies.

\begin{figure}
\begin{center}
\includegraphics[width=0.45\textwidth]{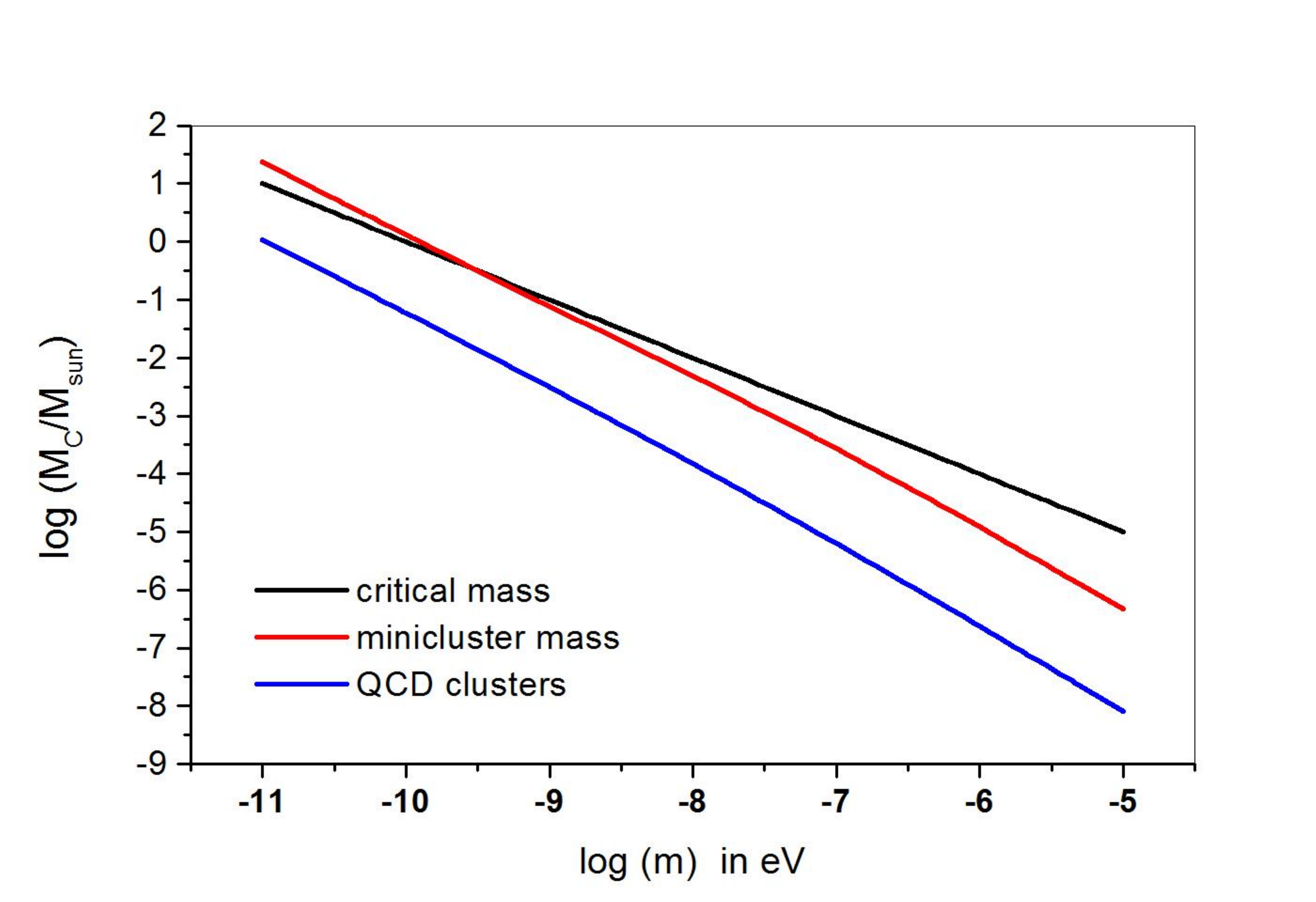} 
\end{center}
\caption{The mass of axionic miniclusters as a function of the axion mass for two different temperature evolution models \cite{fairbairn}. The blue curve
corresponds to QCD axions while the red curve corresponds to a rapid variation of the axion mass with temperature. 
The black curve corresponds to the critical mass derived by \cite{EPJC}.}
\label{fig2}
\end{figure}

As mentioned before, in the scenario where primordial axion miniclusters are formed, the estimated masses are below the critical value and hence they will not
collapse into black holes. The early evolution of axions is determined mainly by two energy scales: the mass $m$ and the decay constant $f_a$. The time $t_0$
at which the axion mass becomes significant is when its Compton wavelength is comparable to the Hubble radius, that is $m \approx hH(t_0)/c^2$. This mass
is a consequence of non-perturbative effects like instantons \cite{gross} and its evolution with the temperature can be modeled by the relation
$m \propto T^{-n}$. For QCD axions $n =3.34$, a value which is consistent with lattice simulations. The computed mass of miniclusters by \cite{fairbairn}
as a function of the QCD axion mass is shown in Fig. \ref{fig2} (blue curve) that is always below the minimum mass (black curve) computed from the relation
derived by \cite{EPJC}.

Higher minicluster masses can be obtained if a more dramatic variation of the axion mass with the temperature is considered. Assuming for instance $n=6$, 
minicluster masses up to $10^3\,M_\odot$ can be obtained. These large values are not in contradiction with constraints imposed by the Lyman-$\alpha$ forest power spectrum
\cite{fairbairn}. The minicluster mass as a function of the axion mass for the case $n=6$ is also shown in Fig. \ref{fig2} (red curve). Inspection of these plots
indicate that, for the temperature evolution scenario with $n=6$, miniclusters with masses above $0.32\,M_\odot$ corresponding to an axion mass
of $3 \times 10^{-10}$ eV are susceptible to undergo the gravitational collapse.

Here it is assumed that a fraction $\eta$ of axion (or ALP) miniclusters of mass $M \sim 0.32\,M_{\odot}$ constituting dark halos has collapsed and
formed black holes. Upper limits from microlensing implies that the maximum contribution of black holes of such a mass to the total dark matter
density is about $\eta \approx 0.05$ (see \cite{freese}), a value that will be adopted in our computations. We assume also that a fraction $f_b$ of
these black holes form binary systems, a parameter that will be constrained by the coalescence rate of these systems as we shall see below.

\section{The evolution of axion black hole binaries}

The putative binaries constituted by two black holes of mass $M_{bh} = 0.32 \, M_\odot$ have an orbital separation distribution 
that fixes the rate at which they merge due to energy and angular momentum losses by gravitational radiation. During the inspiral 
phase of the merger, the wave frequency increases and peaks around the orbital frequency corresponding to a pair separation close to
the gravitational radius. Such a characteristic frequency scales inversely with the black hole mass and is given by \cite{buonnano}
\begin{equation}
f_{\text{max}}=4397\left(1+0.316\gamma\right)\left(\frac{M_\odot}{M}\right) \text{Hz},
\label{fmax}
\end{equation}
where $M$ is the binary mass, which in our case is $0.64\,M_\odot$, and $\gamma = \mu/M$ with $\mu$ being the reduced mass of the binary. Putting
numbers, it results that the maximum gravitational wave frequency is about $7.4$ kHz.

In the absence of a detailed mechanism for the formation of black hole pairs, despite recent investigations in this sense \cite{liu}, only a 
simple estimation of the physical characteristics of the pairs will be presented here.
The main free parameter of our model, the fraction of binaries $f_b$, will be constrained by requiring merger rates respectively equal to 1 event each
ten years, 1 event per year and an optimistic assumption of 10 events per year that would occur in the volume of the universe probed by the gravitational
antenna. This point will be discussed in more detail in the next section.
We will first assume that when dark halos begin to be assembled some $10^{10}$ yr ago, the initial number of binaries $N_0$ in a given halo is approximately 
\begin{equation}
N_0=\eta f_b \left( \frac{M_H}{M} \right).
\end{equation}
In the equation above $\eta$ is the fraction of dark matter under the form of black holes, $f_b$ is the fraction of binaries of mass $M$ 
among these black holes and $M_H \sim 10^{12}\,M_\odot$ is a typical halo mass for galaxies similar to the Milky Way.
As mentioned before, we will take $\eta = 0.05$. 

In a second step, we will assume that the binaries have a distribution of separation $a$ such as $P(a)da$ is the fraction of binaries with separation in
the interval $a, a+da$. Fixing the masses of the pair components, the merger timescale $\tau$ due to gravitational radiation depends only on the initial
separation as $\tau \propto a^4$. Therefore, the probability per unit of time $P(\tau)$ for the occurrence of a merger after the assembly of the halos is
\begin{equation}
P(\tau)=P(a)\mid\frac{da}{d\tau}\mid \propto \frac{P(a(\tau))}{\tau^{3/4}}.
\end{equation} 
In the absence of a detailed formation mechanism for the binaries, we assume that the pair separation distribution is the same as that observed for massive
stars in our galaxy, that is, $P(a) \propto 1/a$ (see, for instance, \cite{renee}), which was also adopted for estimations of the coalescence rate of neutron 
star binaries \cite{PRVS}. In this case, the merger probability per unit of time at the age $\tau$ is
\begin{equation}
P(\tau) = \frac{K}{\tau},
\end{equation}
where $K$ is a normalisation constant that can be computed by integrating $P(\tau)$ in the interval $\tau_0$ and $T_m$. Therefore, 
$K = \left[\ln(T_m/\tau_0)\right]^{-1}$.
The lower limit $\tau_0$ corresponds to the merger timescale related to the minimal pair separation while the upper limit $T_m$ corresponds to the maximum 
pair separation. In order to maintain the dynamical stability of the pairs, the maximum separation cannot be larger than the mean distance between 
black holes when the halo was assembled. In our estimates we will assume that $T_m$ is about 10 times the age of the Galaxy, namely $T_m \sim 10^{11}$ yr, corresponding
to a maximum initial pair separation of about $4.5\times 10^5$ km, and $\tau_0 = 10^6$ yr.

If $N_b(t)$ is the number of binaries at the instant $t$ (computed after the assembly of the halo), the merger rate $R_b(t)$ is given by
\begin{equation}
R_b(t) = N_b(t)P(t).
\end{equation}
Since the number of binaries varies as
\begin{equation}
\frac{dN_b(t)}{dt} = -R_b(t),
\end{equation}
integration of this equation gives the evolution of the number of pairs in the halo, that is
\begin{equation}
N_b(t) = N_0\left(\frac{\tau_0}{t}\right)^K.
\end{equation}
Using this result, the merger rate at the present age $T = 10^{10}$ yr is
\begin{equation}
R_b(T) = K\frac{N_0}{T}\left(\frac{\tau_0}{T}\right)^{K}L(D).
\label{rate}
\end{equation}
The term $L(D)$ introduced in Eq. (\ref{rate}) needs a more detailed explanation. It takes into account the effective
number of halos inside the volume of radius $D$ probed by the gravitational antenna. Here we follow the procedure adopted by \cite{PRVS} that
can be summarised as follows: assuming  that the light distribution of galaxies tracks dark matter, the factor $L(D)$ represents the ratio
between the total (blue) luminosity within the considered volume and that of the Milky Way. For $D < 0.5$ Mpc the correction factor is essentially
unity since the closest bright galaxy (M31) is at a distance of about $0.77$ Mpc. To compute the correction factor for large distances, the counts
of galaxies performed by \cite{courtois} and the LEDA (Lyon-Meudon Extragalactic Database) were adopted. Following \cite{PRVS} we have also
included the contribution of the large complex of galaxies centered in Norma (the Great Attractor), which alone gives a contribution to the
correction factor of 2900 for distances $D > 70$ Mpc. Hence, in other to make numerical estimates from these relations, it is necessary to
evaluate the volume of radius $D$ probed either by the advanced LIGO or the ET interferometers.

\begin{table*}
	\centering
\begin{tabular}{l|c|c|c|c}
\hline
Detector & $D_{\text{max}}$ (Mpc) & $f_b$ (0.1 events/yr) & $f_b$ (1 event/yr) & $f_b$ (10 events/yr)\\ [1ex]
\hline
Einstein Telescope & $430$ & $9.3 \times 10^{-8}$ & $9.3 \times 10^{-7}$ & $9.3 \times 10^{-6}$ \\[1ex]
\hline
 Advanced LIGO & $57$ & $1.1 \times 10^{-5}$ & $1.1 \times 10^{-4}$ & $1.1 \times 10^{-3}$ \\[1ex]
 \hline
\end{tabular}
\caption{Minimum fraction of axionic black hole binaries derived from three assumed detection rates (in brackets) for
the Einstein Telescope and Advance LIGO. The maximum radius of the probed volume of the universe is also given.}
\label{table}
\end{table*}

\section{Detection limits}

The strength of a given signal is characterised by the signal-to-noise ratio ($S/N$), which depends on the
source power spectrum and on the noise spectral density $S_n(f)$ of the detector. For the merging of a
binary system constituted by compact objects, the optimum $S/N$ ratio is obtained by the matched-filtering technique, e.g.,
\begin{equation} \label{S/N}
\left( \frac{S}{N} \right)^2 = 4 \int_0^{\infty} \frac{\tilde{h}^2(f)}{S_n(f)}df.
\end{equation}
In the equation above, $\tilde{h}^2(f)$ is the sum of the square of the Fourier transform of both polarization components of the gravitational wave 
signal. Here we will adopt the approach by Finn \cite{Finn} and, in this case, equation (\ref{S/N}) is reduced to
\begin{equation} \label{S/N2}
\frac{S}{N} = 8 \Theta \left( \frac{r_0}{D} \right) \left( \frac{M_*}{1.2M_{\odot}} \right)^{5/6} \zeta(f_{max}).
\end{equation}
In this equation the angular function $\Theta$ depends on the geometrical projection factors of the detector and on the inclination angle $i$ between the 
orbital angular momentum of the binary and the line of sight, that is,
\begin{equation}
\Theta^2 = 4\left[ \left( 1+\cos^2i \right)^2 F_+^2 + 4 \cos^2i\, F_{\times}^2 \right].
\end{equation}
According to Finn \& Chernoff \cite{Finn2}, if $\Theta$ is in the range $0 \leq \Theta \leq 4$ , its probability
distribution can be approximated with a quite good accuracy by the relation
\begin{equation}
P(\Theta) = \frac{5\Theta(4-\Theta)^3}{256}.
\end{equation}
Using this distribution, the average value of $\Theta$ that will be adopted in our computations is $4/3$. The other quantities in 
equation (\ref{S/N2}) are: the chirp mass $M_* = \mu^{3/5} M^{2/5}$  with $\mu$ and $M$ being respectively the reduced and the total mass of the system, $D$ is the 
distance to the source, and the parameter $r_0$ (having the
dimension of a length) is defined by the relation
\begin{equation}
r_0 = 9.25 \times 10^{-19} \sqrt{I_{7/3}}\; \text{kpc},
\end{equation}
where
\begin{equation} \label{6}
I_{7/3} = \left( \frac{f_0}{\pi} \right)^{1/3} \int_0^{\infty} \frac{df}{f^{7/3}S_n(f)},
\end{equation}
with $f_0 = 203.38$ kHz. The factor $\zeta(f_{\text{max}})$ is defined similarly, e.g.,
\begin{equation} \label{7}
\zeta(f_{\text{max}}) = \frac{(f_0/\pi)^{1/3}}{I_{7/3}} \int_0^{2f_{\text{max}}} \frac{df}{f^{7/3}S_n(f)},
\end{equation}
and the maximum frequency appearing in the upper limit of the integral is given by Eq. (\ref{fmax}). 

In order to compute the integrals in equations (\ref{6}) and (\ref{7}), we adopt the spectral noise density for the advanced LIGO interferometer \cite{pitkin} and 
for the planned Einstein Telescope (ET) \cite{ET}, version shot-noise limited antenna with a knee-frequency around $1$ kHz. 
In these cases the scale parameter results to be $r_0 = 127$ Mpc and $958$ Mpc respectively for the advanced LIGO and ET instruments.
Hence, for $S/N \sim 7$, the typical threshold for a false alarm rate of about one per year, the maximum distances that can be probed respectively
by advanced LIGO and ET are $57$ Mpc and $430$ Mpc.
Using these numbers, the resulting fraction of binaries is given in Table \ref{table} for three assumed detection rates. For having a detection rate of one event per year,
advanced LIGO imposes a limit of about $10^{-4}$ for the fraction of binaries, while a lower limit is needed for the ET antenna that is about
$10^{-6}$.

\section{Concluding remarks}

In the past years, the collapse of a bosonic star has been the subject of many investigations. If several questions are still waiting for an adequate 
answer, it seems that there is a general agreement about the existence of a critical mass above which a black hole could be formed. The original
work by Kaup \cite{Kaup}, based on the solution of the relativistic Klein-Gordon equation, indicated that for the free-field case there is a maximum
mass of the configuration (Kaup limit) that depends inversely on the mass of the scalar field. The equilibrium of these configurations is a consequence of
the balance between gravity and the quantum pressure due to the Heisenberg Uncertainty Principle. More recently, from a dynamical study of scalar 
fields \cite{EPJC}, the authors conclude that a black hole could be formed if the mass of the system is larger than a critical value that differs
from the Kaup limit only by a small numerical factor.

On the other hand, axions (or axion-like particles) are possible candidates to be identified as dark matter particles. Thus, a natural question arises whether these particles
could form structures with masses above the critical value and, consequently, collapse into black holes. As we have seen, QCD axions can form miniclusters
whose mass are below the critical value and hence the formation of black holes is not expected. Nevertheless, a recent investigation \cite{fairbairn}
considered a more extreme scenario in which the axion mass has an important dependence on the temperature, i.e., $m \propto T^{-6}$. In this picture,
for a given axion mass, the corresponding minicluster mass is larger than that of QCD axions. Using this result and the critical mass by \cite{EPJC}, we have
shown that for masses above $0.32\,M_\odot$, corresponding to an axion mass of $3\times 10^{-10}$ eV, the miniclusters can collapse and form black holes. Note that the minicluster model taken from \cite{fairbairn} only guarantees enough mass for the critical condition in the whole cluster, and that we are assuming that the whole cluster is collapsing into the black hole, with no intermediate process of stars formation, that would require an efficient cooling mechanism or an even larger minicluster mass in order to produce a critical axion star \cite{russos2}.

Clearly, it is difficult to say if the adopted scenario for the formation of axion miniclusters is realistic or not. It is worth mentioning that
a recent investigation has simulated the formation of axion stars using a relativistic approach \cite{widdicombe}. These simulations indicate
that configurations with masses larger than $2.14 \times 10^{-10}\, \text{(eV/m)}\,M_\odot$ collapse into a black hole. For axions of mass $3\times 10^{-10}$ eV the
critical mass is about $0.7~M_\odot$, a value that is surprisingly consistent with the values we have obtained.
If in \cite{widdicombe} the possibility of emission of gravitational waves by the coalescence of binaries is invoked, the authors restrict their
analysis to the fact that the maximum emitted frequency will be within the range of frequencies able to be detected by the advanced LIGO interferometer.
In other words, the antenna is able to explore the window of axion masses around $10^{-10}$ eV. 

In this work it is assumed that a fraction of black hole binaries is formed during the process of assembly of halos as described in \cite{fairbairn}
and a simple model for their evolution is presented, which is controlled essentially by the gravitational radiation emitted during the inspiral phase.
Two interferometers were considered in our analysis: advanced LIGO and the planned Einstein Telescope. The first can probe the signal of these
binaries in a volume with radius of $57$ Mpc while the second will see much further, probing a volume having a radius of $430$ Mpc.
For detection rates of one event per year, the limits imposed to the binary fraction are  of $10^{-4}$ and $10^{-6}$ respectively for the advanced LIGO and ET.
In the case of ad-LIGO that is currently in operation, ten years of accumulated data will move an order of magnitude this limit and impose severe constraints either on the formation of binaries or on the fraction of black holes that could have been formed in the axion
dark matter scenario. 

\newpage

\section*{Acknowledgements}

We are thankful to O. Aguiar and R. Sturani for valuable references and discussions, and to CNPq and FAPES for financial support. SC is grateful to P.C. de Holanda and F. Sobreira for the hospitality at UNICAMP.

\end{document}